\documentclass[prx,twocolumn,amsmath,amssymb]{revtex4}

\usepackage{graphicx}
\usepackage{dcolumn}
\usepackage{bm}
\usepackage{mathrsfs}
\usepackage{subfigure}
\usepackage{natbib}
\usepackage{amsmath}
\usepackage{amssymb}
\usepackage{Jonasmacros}
\usepackage{mathptmx}
\usepackage{txfonts}
\usepackage{pifont}

\usepackage[colorlinks,citecolor=blue,linkcolor=Fuchsia]{hyperref}
\usepackage[usenames,dvipsnames,svgnames]{xcolor}

\begin{document}

\title{Chirality Induced Spin Selectivity -- The Role of Electron Correlations}

\author{J. Fransson}
\email{Jonas.Fransson@physics.uu.se}
\affiliation{Department of Physics and Astronomy, Box 516, 75120, Uppsala University, Uppsala, Sweden}

\begin{abstract}
Chirality induced spin selectivity, discovered about two decades ago in helical molecules, is a non-equilibrium effect that emerges from the interplay between geometrical helicity and spin-orbit interactions. Several model Hamiltonians building on this interplay have been proposed and while these can yield spin-polarized transport properties that agrees with experimental observations, they simultaneously depend on unrealistic values of the spin-orbit interaction parameters. It is likely, however, that a common deficit originates from the fact that all these models are uncorrelated, or, single-electron theories. Therefore, chirality induced spin selectivity is, here, addressed using a many-body approach, which allows for non-equilibrium conditions and a systematic treatment of the correlated state. The intrinsic molecular spin-polarization increases by two orders of magnitudes, or more, compared to the corresponding result in the uncorrelated model. In addition, the electronic structure responds to varying external magnetic conditions which, therefore, enables comparisons of the currents provided for different spin-polarizations in one of the (or both) leads between which the molecule is mounted. Using experimentally feasible parameters and room temperature, the obtained normalized difference between such currents may be as large as 5 -- 10 \% for short molecular chains, clearly suggesting the vital importance of including electron correlations when searching for explanations of the phenomenon.
\end{abstract}
\date{\today} 
\maketitle



Chirality induced spin selectivity has recently become a research field on its own merits, and refers back to the seminal work in which the photo current flowing through double stranded DNA molecules, mounted on a Au surface, was shown to be strongly spin-polarized \cite{Science.283.814,Science.331.894}. Due to a combined effect of the helical structure, spin-orbit interaction, and non-equilibrium conditions, the current carrying electrons become spin-polarized. In this fashion, the helical molecular structure acts as a spin-filter. The chirality induced spin selectivity phenomenon has been shown to not be limited to multi-stranded helical structures such as double stranded DNA molecules \cite{NanoLett.11.4652} and bacteriorhodopsin \cite{PNAS.110.14872}, but the effect arises also in, for example, $\alpha$-helicial oligopeptides \cite{NatComm.7.10744,NatComm.8.14567,AdvMat.30.1707390,JPhysChemLett.10.1139} and polyalanines \cite{NatComm.4.2256,NanoLett.14.6042,Small.15.1804557}, and recently also in helicene \cite{AdvMater.28.1957,JPhysChemLett.9.2025}.

On the theoretical side, an enormous effort has been devoted to achieve attractable model descriptions which enable reproduction of the experimental results, in terms of experimentally viable parameters, as well as to generate a sound comprehension of chirality induced spin selectivity. Different approaches can be categorized into hydro-dynamical, or, continuum models \cite{JChemPhys.131.014707,EPL.99.17006,JPCM.26.015008,PhysRevB.88.165409,JChemPhys.142.194308,PhysRevE.98.052221,PhysRevB.99.024418,NJP.20.043055,JPhysChemC.123.17043}, tight-binding descriptions \cite{PhysRevB.85.081404(R),PhysRevLett.108.218102,JPhysChemC.117.13730,PhysRevB.93.075407,PhysRevB.93.155436,ChemPhys.477.61}, and \emph{ab initio} simulations using, for example, density functional theory \cite{JPhysChemLett.9.5453,JPhysChemLett.9.5753,chemrxiv.8325248}.

A necessary condition to generate spin-polarized transport that emerges in all these theoretical reports is the existence of an intrinsic spin-orbit coupling alongside the helical geometry of the molecules. For instance, in Refs. \citenum{JPhysChemLett.9.5453,chemrxiv.8325248}, it is demonstrated that spin-polarized transport cannot be achieved by spin-orbit interaction alone in non-helical molecules, or solely by helicity in the absence of spin-orbit interactions. Moreover, in single stranded helices, a necessary requirement for spin-polarized transport appears to be a finite spin-orbit interaction either between nearest-neighboring ionic sites \cite{PhysRevB.93.075407} or on-site \cite{PhysRevB.85.081404(R)}. While a finite spin-orbit interaction between nearest-neighboring sites may be reasonable when considering, for example, $p$-wave hopping, it yet remains an open question whether the on-site spin-orbit interaction is non-negligible in the types of organic molecules pertaining to spin selectivity. By contrast, in double-, or multi-stranded helices this condition, can be relaxed by introducing inter-helix electron tunneling \cite{PhysRevLett.108.218102}, since there is an inter-helix spin-orbit interaction that may be appreciable.

Nevertheless, while all these theoretical approaches are able to reproduce spin-polarized transport properties, they all rely on unrealistic spin-orbit interaction parameters. While the source of this deficit remains unclear, it is, however, likely to be an effect of the uncorrelated, or, single-electron nature of these models. Thus far, there has been no attempt to address the spin selectivity phenomenon in terms of correlated electrons. In this article, the importance of electron-electron interactions in the modeling of spin-selectivity is demonstrated. To this end, the electronic structure of the helical molecule is described with a model comprising on-site Coulomb interactions, accompanied by nearest-neighbor hopping and next nearest-neighbor spin-orbit interactions. Using experimentally feasible parameters at room temperature, it is shown that the on-set of the local Coulomb interactions may enhance the local molecular spin-polarization by more than two orders of magnitude, and increasing the spin-polarization of the current from negligible (in absence of Coulomb interactions) to a few percents, or more. The spin-polarization originates under non-equilibrium from the cooperation of the helical geometry and the spin-orbit interaction. However, in absence of electronic correlations, there is no spin exchange introduced and, hence, the spin-polarization of the electronic ground state remains negligible. By contrast, finite on-site Coulomb interactions do provide the necessary exchange between the spin channels, already at the mean field level, which allows the spontaneously broken spin symmetry to become finite and viable for spin-polarized transport.



\begin{figure}[t]
\begin{center}
\includegraphics[width=\columnwidth]{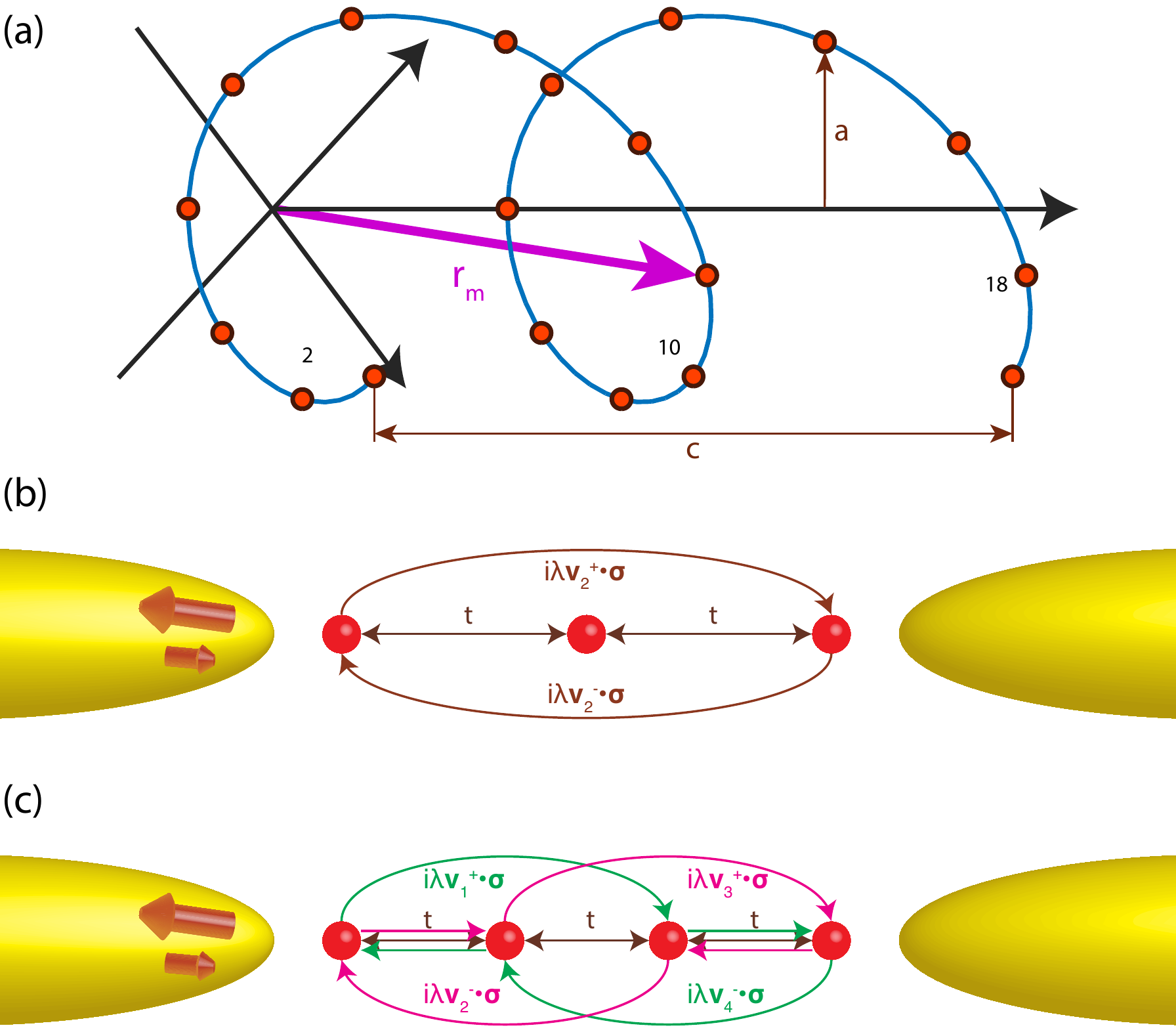}
\end{center}
\caption{(a) Sketch of the helical molecule considered in this text. In our convention, this molecule has negative chirality (counterclockwise helicity). (b), (c), Simplified layouts of molecules with (b) three and (c) four ionic sites, illustrating possible hopping and spin-orbit interaction paths. The sketch shows how the molecules are attached to a left ferromagnetic lead with majority spin up \emph{or} down, and a right non-magnetic lead.}
\label{fig-Fig1}
\end{figure}

The generic molecular geometry considered in this article is described by the set of spatial coordinates
\begin{align}
\bfr_m=&
	\Bigl(a\cos\varphi_m,a\sin\varphi_m,(m-1)c/(MN-1)\Bigr)
	,
\\&
	\varphi=(m-1)2\pi/N
	,\
	m=1,\ldots,MN
	,
\nonumber
\end{align}
where $a$ and $c$ define the radius and length, respectively, of the helical structure, as is illustrated in Fig. \ref{fig-Fig1} (a), whereas $M$ and $N$ denote the number of laps and ions per lap, such that $\mathbb{M}=MN$ is the total number of sites. Each coordinate denotes an ionic site which is represented by a single electron level described by $\dote{m}\psi^\dagger_m\psi_m+U_mn_{m\up}n_{m\down}$, where $\psi_m^\dagger=(\ddagger{m\up}\ \ddagger{m\down})$ ($\psi_m$) is the creation (annihilation) spinor, $\dote{m}$ denotes the energy level at the site, and $U_m$ is the electron-electron interaction energy, whereas $n_{m\sigma}=\ddagger{m\sigma}\dc{m\sigma}$, $\sigma=\up,\down$, is the electron number operator. Electron hopping between nearest-neighboring sites occurs with the energy $t$, and spin-orbit coupling is picked up between next-nearest neighbor sites through processes of the type $i\lambda\psi_m^\dagger\bfv_m^{(s)}\cdot\bfsigma\psi_{m+2s}$, $s=\pm1$, where $\lambda$ denotes the spin-orbit interaction parameter. The vector $\bfv_m^{(s)}=\hat\bfd_{m+s}\times\hat\bfd_{m+2s}$ defines the chirality of the helical molecule in terms of the unit vectors $\hat{\bfd}_{m+s}=(\bfr_m-\bfr_{m+s})/|\bfr_m-\bfr_{m+s}|$; here, positive chirality corresponds to right handed helicity. Finally, $\bfsigma$ denotes the vector of Pauli matrices.
The chiral molecule comprising $\mathbb{M}$ sites can, thus, be modeled by the Hamiltonian
\begin{align}
\Hamil_\text{mol}=&
	\sum_{m=1}^{\mathbb{M}}
		\left(
			\dote{m}\psi^\dagger_m\psi_m
			+
			U_mn_{m\up}n_{m\down}
		\right)
	-
	t
	\sum_{m=1}^{\mathbb{M}-1}
		\Bigl(
			\psi^\dagger_m\psi_{m+1}
			+
			H.c.
		\Bigr)
\nonumber\\&
	+
	\lambda
	\sum_{m=1}^{\mathbb{M}-2}
		\Bigl(
			i\psi^\dagger_m\bfv_m^{(+)}\cdot\bfsigma\psi_{m+2}
			+
			H.c.
		\Bigr)
	.
\end{align}
This model is a generalization of the Kane and Mele model \cite{PhysRevLett.61.2015,PhysRevLett.95.146802,PhysRevLett.95.226801}, including the spin-orbit interaction in all spatial directions and confined to a finite helical molecular structure.

The molecule is coupled to metallic leads by tunneling interactions $\Hamil_T=\sum_\bfp t_\bfp\psi^\dagger_\bfp\psi_1+\sum_\bfq t_\bfq\psi^\dagger_\bfq\psi_N+H.c.$, where the leads are modeled by $\Hamil_\chi=\sum_{\bfk\in\chi}\leade{\bfk}\psi^\dagger_\bfk\psi_\bfk$, $\leade{\bfk}=\dote{\bfk}+\Delta_\chi/2$. Here, $\chi=L,R$ denotes the left ($L$, $\bfk=\bfp$) or right ($R$, $\bfk=\bfq$) lead, and $\Delta_\chi$ denotes the spin-gap in the lead. Hence, the full metal-molecule-metal junction is described by
\begin{align}
\Hamil=&
	\Hamil_L+\Hamil_R+\Hamil_\text{mol}+\Hamil_T
	.
\end{align}

The properties of the electronic structure, for example, the density of electron states $\rho_m(\omega)$ and spin moment $\av{\bfS_m}$, are connected to the single electron Green functions $\bfG_{mn}(\omega)=\av{\inner{\psi_m}{\psi^\dagger_n}}(\omega)$, through $\rho_m(\omega)=i{\rm sp}[\bfG^>_{mm}(\omega)-\bfG^<_{mm}(\omega)]/2\pi$ and $\av{\bfS_m}=(-i){\rm sp}\int\bfsigma\bfG^<_{mm}(\omega)d\omega/(4\pi)$, where ${\rm sp}$ denotes the trace over spin 1/2 space. Here, all quantities are calculated using non-equilibrium Green functions $\bfG^{</>}_{mn}(\omega)$, for the sake of capturing non-equilibrium properties of the system.

Concerning the Green functions, two different approaches are employed, here referred to as uncorrelated and correlated approaches. First in the uncorrelated theory ($U=0$), the Green functions $\bfG_{mn}$ are given in terms of $2\times2$-matrices spanning over spin 1/2 space. The corresponding equations of motion can generally by written as $\bfg^{(-1)}_m\bfG_{mm'}=\delta_{mm'}\sigma^0+\bfSigma_{mn}\bfG_{nm'}$, where the bare Green function $\bfg_m=g_m\sigma^0$ in absence of any predefined spin-polarization, $\sigma^0$ denotes the two-dimensional unit matrix, and where summation over repeated indices is understood.

In the correlated theory, however, the local Hilbert space is expanded to a four dimensional space due to the expansion $\dc{m\sigma}=\X{0\sigma}{m}+\eta_\sigma\X{\bar\sigma2}{m}$ \cite{PhysRevB.72.075314,NonEquilibriumNanoPhysics}, where the Hubbard operator $\X{pq}{}\equiv\ket{p}\bra{q}$ describes a transition from the (many-body) state $\ket{q}$ to the state $\ket{p}$. As a result of the algebra that accompanies the Hubbard operators $\X{a}{m}$, the equations of motion for the Green function become $\bfd^{-1}_m\bfG_{mm'}=\bfP_{mm'}+\bfP_{mm}\bfSigma_{mn}\bfG_{nm'}$, where a bare Green function $\bfg_m$ is defined by $\bfg_m=\bfd_m\bfP_{mm}$ and where $\bfP_{mn}=\delta_{mn}\{\av{\anticom{\X{a}{m}}{\X{\bar{b}}{m}}}\}_{a\bar{b}}$ provides spectral weights to the transitions involved. A critical difference between the uncorrelated and correlated approaches is that, while the bare Green functions $\bfg_m$ commute in the former approach they \emph{do not} in the latter.
This difference in commutation properties can be traced back to the Coulomb interaction, where the non-commutativity is a manifestation of the exchange interaction introduced by the local Coulomb repulsion. The consequences of this property is discussed in more detail below.
In the present context, the Green functions are provided at the mean field level, Hubbard I approximation \cite{PhysRevB.72.075314,NonEquilibriumNanoPhysics}, which is justified for room temperature calculations.



In the present model, a chain of at most three sites cannot generate a mechanism for spin symmetry breaking. This can, for instance, be seen through considering the equation of motion for the Green function at site $m=1$, which to the third order scattering processes is given by
\begin{align}
\bfG_1^{-1}=&
		\bfg^{-1}_1
		-
		t\bfg_2t
		-
		t\bfg_2t\bfg_3i\lambda\bfv_3^{(-)}\cdot\bfsigma
\nonumber\\&
		-
		i\lambda\bfv_1^{(+)}\cdot\bfsigma\bfg_3t\bfg_2t
		-
		i\lambda\bfv_1^{(+)}\cdot\bfsigma\bfg_3i\lambda\bfv_3^{(-)}\cdot\bfsigma
	.
\label{eq-G1trimer}
\end{align}
Possible scattering paths are indicated in Fig. \ref{fig-Fig1} (b). In the trimer chain, the chirality vector $\bfv_3^{(-)}\sim(\bfr_3-\bfr_2)\times(\bfr_2-\bfr_1)=-(\bfr_1-\bfr_2)\times(\bfr_2-\bfr_3)\sim\bfv_1^{(+)}$. Hence, in the uncorrelated theory, the third and fourth contributions on the right hand side of Eq. (\ref{eq-G1trimer}) exactly cancel each other. Furthermore, the last contribution becomes spin-independent. By the same argument, any higher order scattering contribution is spin symmetric which implies that the uncorrelated theory for a trimer is, by necessity, spin degenerate. Because of the absence of a spin symmetry breaking mechanism, spin degeneracy is maintained also for finite electron correlation energy, $U>0$. 

The conclusion about the preserved spin symmetry is completely changed for chains with more than three sites. The reason is that molecules comprising four ionic sites or more, allow for scattering processes of the fourth order, or higher, in which the spin-orbit contributions are not necessarily parallel and which, therefore, open up the possibility for spontaneous spin symmetry breaking.
Considering a four site molecule and scattering processes pertaining to the spin degrees of freedom, the two fourth order scattering paths that contribute to the spin symmetry breaking are depicted in Fig. \ref{fig-Fig1} (c). The chiral geometry of the molecular structure implies that the spin orbit interactions coming into play in these scattering paths, through the chirality vectors $\bfv_m^{(s)}$, are inequivalent. To see this, consider the equation of motion for the Green function at site $m=1$, which contains the fourth order processes
\begin{align}
&\bfG_1:
\nonumber\\&
	\bfg_1i\lambda\bfv_1^{(+)}\cdot\bfsigma\bfg_3t\bfg_4i\lambda\bfv_4^{(-)}\cdot\bfsigma\bfg_2t
	+
	\bfg_1t\bfg_2i\lambda\bfv_2^{(+)}\cdot\bfsigma\bfg_4t\bfg_3\lambda\bfv_3^{(-)}\cdot\bfsigma
\nonumber\\&=
	-t^2\lambda^2g_1g_2g_3g_4
	\biggl(
		\Bigl(\bfv_1^{(+)}\cdot\bfsigma\Bigr)\Bigl(\bfv_4^{(-)}\cdot\bfsigma\Bigr)
		+
		\Bigl(\bfv_2^{(+)}\cdot\bfsigma\Bigr)\Bigl(\bfv_3^{(-)}\cdot\bfsigma\Bigr)
	\biggr)
	.
\label{eq-G1quadrimer}
\end{align}
Here, the last equality is obtained by recalling that the bare Green functions $\bfg_m$ and $\bfg_n$ commute for all $m, n$.
It is important to notice that the involved chirality vectors $\bfv_1^{(+)}$ ($\bfv_2^{(+)}$) and $\bfv_4^{(-)}$ ($\bfv_3^{(-)}$) are not parallel with one another. As an effect, the product, for example, $(\bfv_1^{(+)}\cdot\bfsigma)(\bfv_4^{(-)}\cdot\bfsigma)=\bfv_1^{(+)}\cdot\bfv_4^{(-)}+i(\bfv_1^{(+)}\times\bfv_4^{(-)})\cdot\bfsigma$, comprises a cross product which is finite. This non-vanishing spin-dependent component does indeed open up for a mechanism for spontaneous spin symmetry breaking.

Before proceeding it is worth to notice that, the fourth order scattering processes also involve paths like, for example, \ding{172}$\rightarrow$\ding{174}$\rightarrow$\ding{175}$\rightarrow$\ding{174}$\rightarrow$\ding{172} and \ding{172}$\rightarrow$\ding{173}$\rightarrow$\ding{175}$\rightarrow$\ding{173}$\rightarrow$\ding{172}, mathematically expressed as 
$\bfg_1i\lambda\bfv_1^{(+)}\cdot\bfsigma\bfg_3t\bfg_4t\bfg_3i\lambda\bfv_3^{(-)}\cdot\bfsigma$ and $\bfg_1t\bfg_2i\lambda\bfv_1^{(+)}\cdot\bfsigma\bfg_4i\lambda\bfv_4^{(-)}\cdot\bfsigma\bfg_2t$, respectively. However, since the chirality vectors $\bfv_1^{(+)}$ ($\bfv_2^{(+)}$) and $\bfv_3^{(-)}$ ($\bfv_4^{(-)}$) are parallel with each other, the spin-dependent component vanishes identically, hence, those processes do not contribute to the spin symmetry breaking mechanism.

While crucial for the emergence of a spin-polarization, the chirality and spin-orbit interactions alone is not enough for the structure to sustain an appreciable spin-polarization. The reason for this inability can be traced back to the spin symmetry breaking mechanism itself. For an order of magnitude estimate, consider the first contribution in Eq. (\ref{eq-G1quadrimer}) in the uncorrelated picture, for example, which can be rewritten as
\begin{align}
-
	t^2\lambda^2
	g_1^4
	\Bigl(
		\bfv_1^{(+)}\cdot\bfv_4^{(-)}
		+
		i[\bfv_1^{(+)}\times\bfv_4^{(-)}]\cdot\bfsigma
	\Bigr)
	,
\label{eq-SEfact}
\end{align}
where, without loss of generality, the bare Green functions $g_m$ have been replaced by $g_1$ for all $m$. The energy scale associated with the spin-polarization generated by the fourth order scattering processes is, therefore, set by two times the largest of $t^2\lambda^2g_1^4|\bfv_1^{(+)}\times\bfv_4^{(-)}|$ and $t^2\lambda^2g_1^4|\bfv_2^{(+)}\times\bfv_3^{(-)}|$. Clearly, the energy scale is second order in both the hopping and spin-orbit interaction parameters and, hence, the induced spin-polarization is necessarily small in the uncorrelated picture. This conclusion does not change by summing to all orders of the analogous contributions in the higher order scattering processes.

\begin{figure}[t]
\begin{center}
\includegraphics[width=\columnwidth]{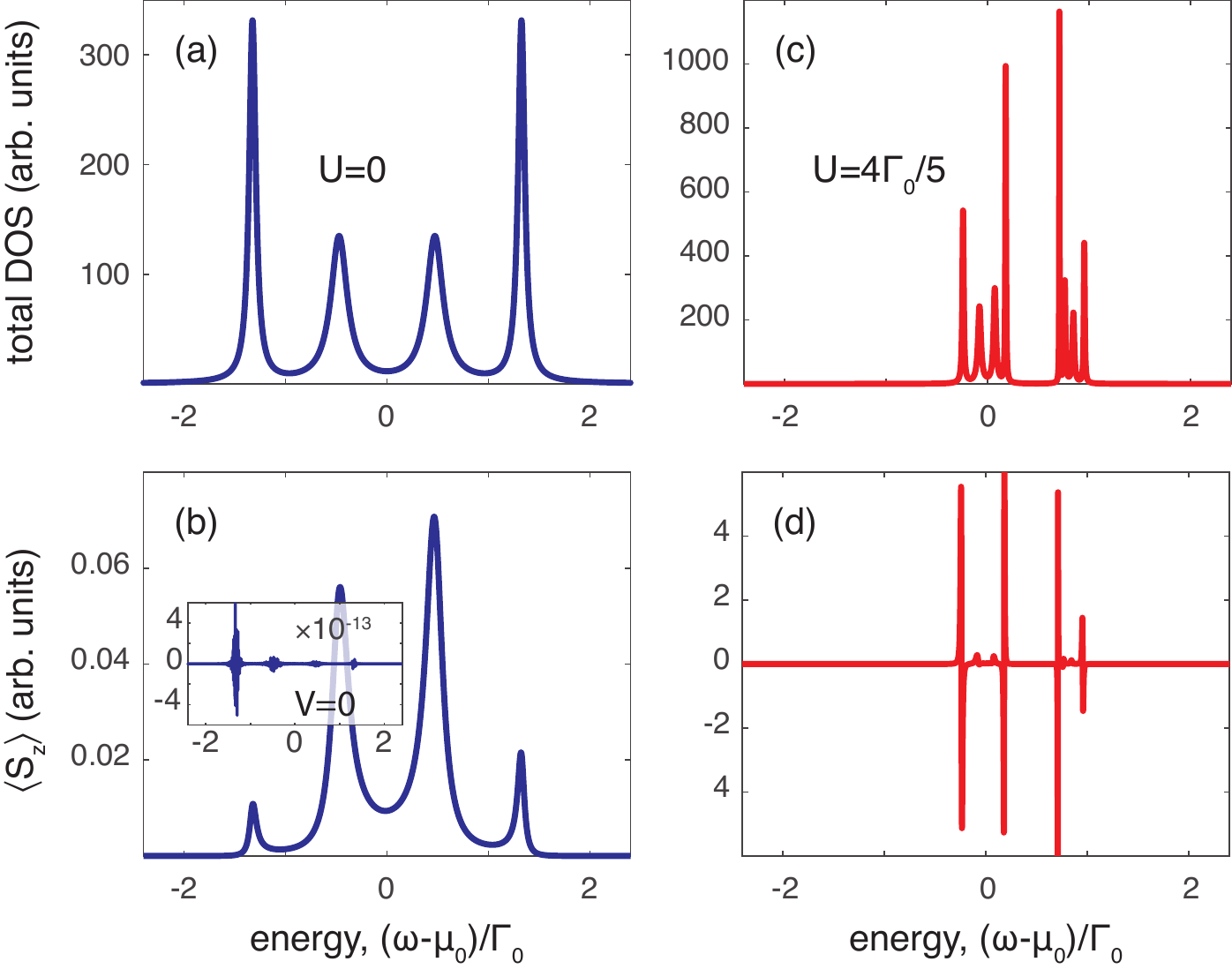}
\end{center}
\caption{Non-equilibrium (a), (c) electronic and (b), (d) spin-polarization of a four site molecule in (a), (b) the uncorrelated ($U=0$) and (c), (d) correlated ($U=4\Gamma_0/5$) models. The inset in panel (b) shows the equilibrium spin-polarization. Here, $\Gamma_0=50$ meV, $\Gamma^R=10\Gamma^L=\Gamma_0$, $\mu-\mu_0=0$, $V=\Gamma_0/5e$, $t=4\Gamma_0/5$, $\lambda=\Gamma_0/5$, and $T\sim\Gamma_0/2k_B$ (300 K).}
\label{fig-Intrinsic}
\end{figure}

The conclusion drawn above is corroborated by the exact numerical solution obtained through matrix inversion. The plots in Fig. \ref{fig-Intrinsic} (a), (b), show the calculated non-equilibrium total (a) density of electron states (DOS) and (b) spin-polarization $\av{S_z}(\omega)$ for a molecule with four sites coupled to non-magnetic leads and with a finite bias voltage applied across the junction. In absence of electronic correlations, the spectrum is governed by hopping and spin-orbit interactions, which renders a symmetric density of electrons states around the molecular electro-chemical potential $\dote{m}=\dote{0}$, for all $m$. A finite spin-polarization is induced in the molecule, as is illustrated in Fig. \ref{fig-Intrinsic} (b), which demonstrates that the spin symmetry spontaneously breaks due to the cooperation of the geometrical helicity and the spin-orbing interactions. The inset shows that the equilibrium ($V=0$) spin-polarization is negligible, suggesting that spin symmetry breaking becomes effective only under non-equilibrium conditions, in agreement with experimental observations.
In these calculations, the energy scale is set by $\Gamma_0=50$ meV and the couplings to the leads $\Gamma^R=10\Gamma^L=\Gamma_0$, $\dote{0}-\mu=0$, where $\mu$ is the overall chemical potential for the system as a whole and the bias voltage, $V=\Gamma_0/5e$. Moreover, $t=4\Gamma_0/5$, and $\lambda=\Gamma_0/5$, whereas the temperature $T\sim\Gamma_0/2k_B$ (300 K).

By contrast, electrons correlations, that is, finite electron-electron interactions $U$, changes this conclusion quite dramatically. As a way to understand the essential difference with the uncorrelated picture, again consider the first contribution in Eq. (\ref{eq-G1quadrimer}). In the many-body approach employed here, this contribution has to be rewritten according to
\begin{align}
\bfg_1i\lambda\bfV_1^{(+)}\bfg_3\bft\bfg_4i\lambda\bfV_4^{(-)}\bfg_2\bft
	,
\end{align}
where the matrices $\bft=t(\sigma^0\otimes\sigma^0+\sigma^x\otimes\sigma^z)$ and
\begin{align}
\bfV_m^{(s)}=&
	\begin{pmatrix}
		\bfv_m^{(s)}\cdot\bfsigma & \bfv_m^{(s)}\cdot\bfsigma\sigma^z \\
		\sigma^z\bfv_m^{(s)}\cdot\bfsigma & \sigma^z\bfv_m^{(s)}\cdot\bfsigma\sigma^z
	\end{pmatrix}
	.
\end{align}
This is required because of the expansion of the Hilbert space associated with spin 1/2 to a four dimensional space. Moreover, in this framework the bare Green function $\bfg_m$ is the product of two matrices, that is, $\bfd_m\bfP_m$, which do not commute ($\com{\bfd_m}{\bfP_m}\neq0$), in general, as do not the bare Green functions of different sites. The non-commutativity is a general feature of any many-body approach in which electron correlations are included, and is, here, a signature of the exchange interaction introduced between the spin channels.

The exchange interaction does not by itself provide a mechanism for breaking the spin symmetry, something which can be tested by removing either the helicity and/or the spin-orbit interactions from the calculations. However, since the spin symmetry breaking mechanism, introduced by the helicity and spin-orbit interactions, is an intrinsic property of the system, the exchange interaction provided through the local Coulomb repulsion, leads to distinct spin-dependent spectral weights, for example, $P_m^{0\up}\neq P_m^{0\down}$, where $P_m^{0\sigma}\equiv\av{\anticom{\X{0\sigma}{m}}{\X{\sigma0}{m}}}$. The spectral weights have to be determined self-consistently through the equations
\begin{subequations}
\begin{align}
P_m^{(0\sigma)}=&
	\frac{i}{2\pi}
	\int
		\biggl(
			\sum_{\sigma'}
				\Bigl(\bfG_m^>(\omega)\Bigr)_{0\sigma'}
			-
			\Bigl(\bfG_m^<(\omega)\Bigr)_{0\sigma}
		\biggr)
	d\omega
	,
\\
P_m^{(\sigma2)}=&
	\frac{i}{2\pi}
	\int
		\biggl(
			\Bigl(\bfG_m^>(\omega)\Bigr)_{\sigma2}
			-
			\sum_{\sigma'}
				\Bigl(\bfG_m^<(\omega)\Bigr)_{\sigma'2}
		\biggr)
	d\omega
	,
\end{align}
\end{subequations}
where the subscripts on the parentheses signifies which matrix element to be integrated. Conservation of probability requires that, for example, $\sum_m(P_m^{(0\up)}+P_m^{(\down2)})=1$. These self-consistent equations are non-linear in the spectral weights, which may lead to substantial enhancements of the spin-polarization.

The plots in Fig. \ref{fig-Intrinsic} (c), (d), display the self-consistently obtained non-equilibrium total (c) density of electron states and (d) spin-polarization, calculated under the same conditions as in the uncorrelated case in panels (a), (b), however, with a finite on-site Coulomb repulsion ($U=4\Gamma_0/5$). First, one notices that the electronic density splits up into two portions and that the overall width is reduced, which are expected results for correlated electrons. Second, the peaks of the spin-polarization are about two-orders of magnitude larger than in the uncorrelated case, which is attributed to the introduction of local exchange interaction.

Having demonstrated the influence of the exchange on the spectral properties, the next step is to consider the accompanied transport properties. Experimentally, the current spin-polarization, ${\rm SP}=(J_\up-J_\down)/(J_\up+J_\down)$, can be recorded by, for example, letting one of the leads be ferromagnetic while the other one is non-magnetic \cite{AdvMater.28.1957}. Then, $J_\sigma$ is the charge current measured with majority spin $\sigma$ in the ferromagnetic (left) lead and is, here, calculated through the formula \cite{PhysRevB.72.075314}
\begin{align}
J_\sigma=&
	\frac{ie}{h}
	{\rm sp}
	\int
		\bfGamma^L(\sigma)
		\Bigl(
			f_L(\omega)\bfG_1^>([\bfGamma^L(\sigma)];\omega)
\nonumber\\&\hspace{2cm}
			+
			f_L(-\omega)\bfG_1^<([\bfGamma^L(\sigma)];\omega)
		\Bigr)
	d\omega
	,
\end{align}
where $\bfG_1^{</>}([\bfGamma^L(\sigma)];\omega)$ indicates that the calculated electronic properties depend on the spin-polarization in the left lead through the coupling parameter $\bfGamma^L(\sigma)$. The coupling parameter $\bfGamma^\chi=\Gamma^\chi(\sigma^0+p_\chi\sigma^z)$ defines the effective spin-resolved tunneling rate between the lead $\chi$ and the adjacent molecule, where $p_\chi\in[-1,1]$ denotes the spin-polarization of the tunneling rate, whereas $\Gamma^\chi=\pi\sum_{\bfk\in\chi}|t_\bfk|^2\rho_\chi(\dote{\bfk})$ accounts for the roles of the tunneling rate $t_\bfk$ and the density of electron states $\rho_\chi(\dote{\bfk})$ in the lead $\chi$.

\begin{figure}[t]
\begin{center}
\includegraphics[width=\columnwidth]{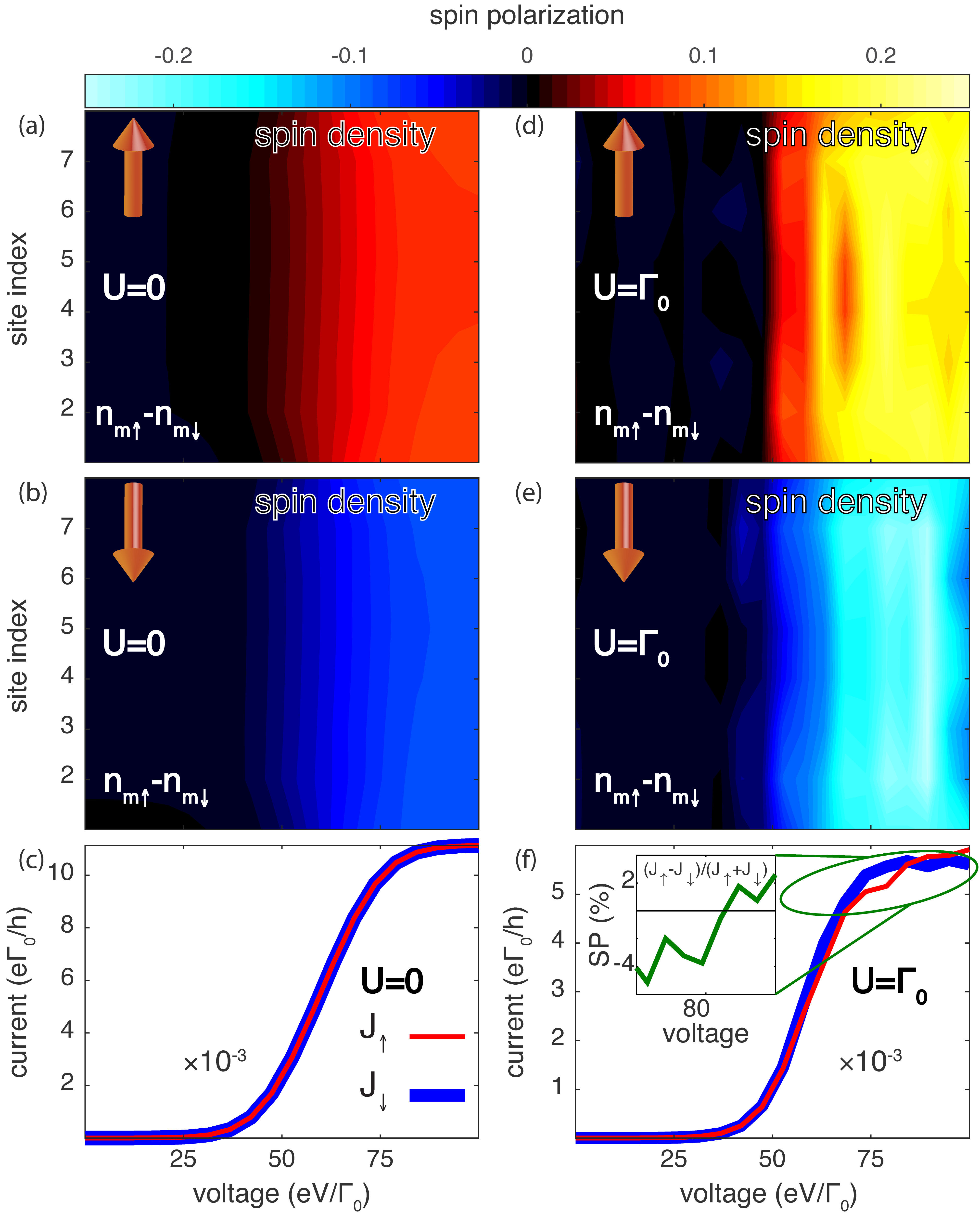}
\end{center}
\caption{(a), (b), (d), (e) Non-equilibrium spin densities of a molecule with eight ionic site mounted in the junction between a ferromagnetic and non-magnetic leads. Panels (a), (b) show results from the uncorrelated ($U=0$) theory whereas panels (c), (d) show the corresponding data from the correlated ($U=\Gamma_0$) theory. The majority spin in the ferromagnetic lead is (a), (d) up and (b), (e), down. (c), (f) Current-voltage characteristics for majority spin up (bold, black) and majority spin down (faint, magenta) in the (c) uncorrelated and (f) correlated theories. The inset in panel (f) shows the current spin-polarization $(J_\up-J_\down)/(J_\up+J_\down)$. Here, $\Gamma_0=10$ meV, $\Gamma^R=10\Gamma^L=\Gamma_0$, and $\Gamma^L=\sum_\sigma\Gamma^L_\sigma$ with $\Gamma^L_\sigma=\Gamma^L(1+\sigma^z_{\sigma\sigma}p_L)/2$ and $p_L=0.5$, whereas $\mu-\mu_0=-30\Gamma_0$, $t=4\Gamma_0$, $\lambda=\Gamma_0/2$, and $T\sim5\Gamma_0/2k_B$ (300 K).}
\label{fig-EightSites}
\end{figure}

The plots in Fig. \ref{fig-EightSites} (a), (b), (d), (e), show the non-equilibrium spin densities resolved per site for a molecule with eight ionic sites (with energy levels $\dote{m}-\mu_0=-30\Gamma_0$, for all $m$), as function of the voltage bias applied across the molecule, where panels (a), (d) [(b), (e)], refer to majority spin $\up$ ($\down$) in the left lead.
In the uncorrelated case, panels (a) and (b), the spin-densities in the two situations conspicuously point in opposite directions which, nonetheless, is an expected result caused by the opposite spin-polarizations of the left lead. Importantly, however, is that the two different conditions render the same degree of spin-polarization, which is a clear manifestation of the essentially negligible intrinsic spin-polarization of the chiral molecule. It is, therefore, expected that the currents resulting from the two different conditions should be nearly equal, which is also shown in Fig. \ref{fig-EightSites} (c). The difference between the two currents is of the order of $10^{-16}$, which should be considered as a numerical error. The conclusion do be drawn out of these results is that the uncorrelated theory (used here) is not capable of capturing a measurable current spin-polarization. By introducing either on-site and/or nearest-neighbor spin-orbit interactions, this conclusion can be revoked \cite{PhysRevB.85.081404(R),PhysRevLett.108.218102}, however, the energy scale of the induced spin-polarization remains to be a power law of the spin-orbit interaction parameters and can, therefore, not viably explain the large effect observed in experiments.

The situation is markedly changed in the correlated case, panels (d) and (e), where both the amplitudes and voltage dependences alter significantly between the two spin-polarizations of the leads. Indeed, depending on the majority spin in the left lead, this injected spin-polarization either competes with, panel (d), or reinforces, panel (e), the intrinsic spin-polarization of the molecule. Under the latter conditions, there is an unstable regime (voltages between 50 and 75) in which the spin-density fluctuates violently between high and low values, while the amplitude of the spin-density increases smoothly in the same regime under the former conditions.

As for the transport properties given by means of the many-body approach, the current through the elastic channel is expected to be maximal whenever the spin channels are degenerate, since the probability to tunnel through the junctions is equally distributed between the spins \cite{PhysRevB.72.045415,PhysRevB.72.075314}. Deviations from degenerate conditions tend to decrease the total current. Here, the non-equilibrium electronic structure is strongly spin-polarized in both situations however, the latter spin-polarization is clearly stronger and also more homogeneously distributed among the ionic sites.
The homogenous spin-polarization creates a more efficient effective (tunneling) barrier between the leads for one of the spin channels.
Therefore, in the present situation, it would be expected that the current in the former case, panel (d), should generate a larger current than the latter, panel (e). The plot in Fig. \ref{fig-EightSites} (f) verifies this conjecture, in particular, at high voltages where the difference between the spin-polarizations is the largest. The inset shows the corresponding spin-polarization of the current, which rises up to more than 4 \%. This should be compared to the essentially vanishing spin-polarization of the current in the uncorrelated approach.

\begin{figure}[t]
\begin{center}
\includegraphics[width=\columnwidth]{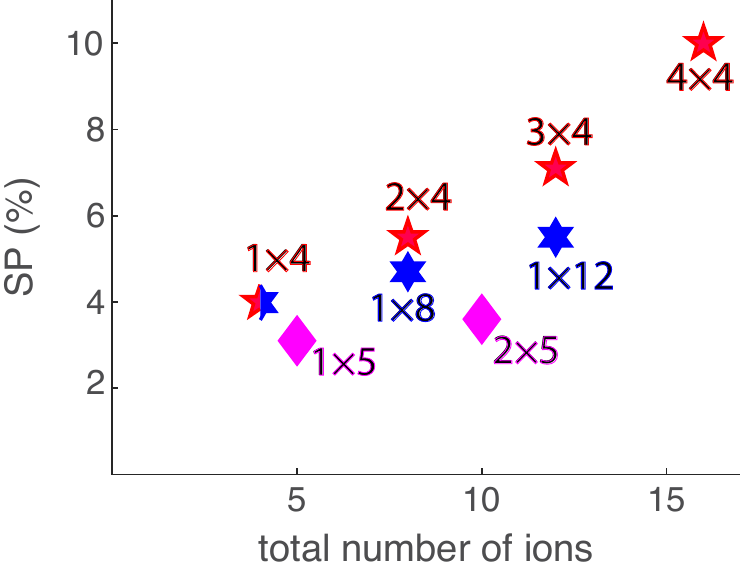}
\end{center}
\caption{Maximum spin polarization of the current as function of the number of ions in the molecular chain. Pentagrams indicate results for molecules with $M=1,\ldots,4$, number of laps with 4 ions per lap, diamonds $m=1,2$, laps with 5 ions per lap, and hexagrams a single lap with 8 and 12 ions. Other parameters are as in Fig. \ref{fig-EightSites}.}
\label{fig-SPtrend}
\end{figure}

While the results presented here are obtained for short chains, up to eight ionic sites, simulations with longer chains show that the maximum spin-polarization of the current increases with the length of the helix. Indeed, computations of chains with $M\times4$, $M=1,2,3$, $M\times5$, $M=1,2$, and $1\times N$, $N=8,12$, see Fig. \ref{fig-SPtrend}, clearly suggest that the spin-polarization of the current should increase with increasing number of laps as well as total number of ions in the molecular chain. The sign change of the spin-polarization seen in the inset of Fig. \ref{fig-EightSites} (f) is most likely a finite size effect and is not considered significant, since it does not appear consistently for increasing number of sites in the chain.

In conclusion, considering helical molecules it has been shown that the spin symmetry is spontaneously broken under non-equilibrium conditions by an interplay of the geometrical helicity and spin-orbit interactions. However, in any single-electron theory, the energy associated with the intrinsically emerging spin-polarization scales as a power law in the spin-orbit interaction. Therefore, it is argued that a theory without electronic correlation cannot explain the large degree of spin-polarization of the transport properties observed in experiments, unless unrealistic values of the spin-orbit interactions are employed. By contrast, using many-body theory and a model with local electron-electron interactions, it has been shown that the degree of spin-polarization is enhanced in all spectral and transport properties considered. Indeed, the local electron-electron interactions give rise to an exchange interaction between the spin channels, which reinforces the chirality induced spin-polarization by two orders of magnitude or more. In addition, the correlated electronic structure responds to varied external magnetic conditions, yielding current spin-polarizations in the order of several percents for short molecular chains.

The results presented here are obtained for short molecular chains and should be considered as a proof of principle. Nonetheless, simulations of the spin-polarized current for longer molecular chains, both with more laps and more total number of ionic sites comprised in the molecule, suggest that the presented theory is capable to account also for influences from the molecular length and periodicity. The amplitude of the spin-polarization of the current increases as function of both the number of ionic sites as well as the number of laps, in good agreement with previous results.

It is likely that additional, more realistic, modeling will further enhance the accuracy of the theory. In particular, the inter-orbital Coulomb interactions that may be included in structures comprising $p$-like orbitals, is expected to introduce an even larger exchange between the spin channels which, then, should give rise to further enhancements of the spin-polarization of the transport properties.

\acknowledgements
The author thanks R. Naaman, L. Nordstr\"om, and A. Sisman for critical comments and advise.
Support from Vetenskapsr\aa det, Stiftelsen Olle Engkvist Byggm\"astare, and Carl Tryggers Stiftelse is acknowledged.


\begin{thebibliography}{200}
\bibitem{Science.283.814} Ray, K.; Ananthavel, S. P.; Waldeck, D. H.; Naaman, R., Asymmetric Scattering of Polarized Electrons by Organized Organic Films of Chiral Molecules, \emph{Science}, {\bf 1999}, \emph{283}, 814.
\bibitem{Science.331.894} G\"ohler, B.; Hamelbeck, V.; Markus, T. Z.; Kettner, M.; Hanne, G. F.; Vager, Z.; Naaman, R.; Zacharias, H., Spin Selectivity in Electron Transmission Through Self-Assembled Monolayers of Double-Stranded DNA, \emph{Science}, {\bf 2010}, \emph{331}, 894--897

\bibitem{NanoLett.11.4652} Xie, Z.; Markus, T. Z.; Cohen, S. R.; Vager, Z.; Gutierrez, R.; Naaman, R., Spin Specific Electron Conduction through DNA Oligomers, \emph{Nano Lett.}, {\bf 2011}, \emph{11}, 4652--4655.
\bibitem{PNAS.110.14872} Mishra, D.; Markus, T. Z.; Naaman, R.; Kettner, M.; G\"ohler, B.; Zacharias, H.; Friedman, N.; Sheves, M.; Fontanesi, C., Spin-dependent electron transmission through bacteriorhodopsin embedded in purple membrane, \emph{Proc. Natl. Acad. Soc.}, {\bf 2013}, \emph{110}, 14872--14876.

\bibitem{NatComm.7.10744} Eckshtain-Levi, M.; Capua, E.; Refaely-Abramson, S.; Sarkar, S.; Gavrilov, Y.; Mathew, S. P.; Paltiel, Y.; Levy, Y.; Kronik, L.; Naaman, R., Cold denaturation induces inversion of dipole and spin transfer in chiral peptide monolayers, \emph{Nat. Comms.}, {\bf 2016}, \emph{7}, 10744.
\bibitem{NatComm.8.14567} Ben Dor, O; Yochelis, S.; Radko, A; Vankayala, K.; Capua, E.; Capua, A.; Yang, S. -H.; Baczewski, L. T.; Parkin, S. S. P.; Naaman, R.; Paltiel, Y., Magnetization switching in ferromagnets by adsorbed chiral molecules without current or external magnetic field, \emph{Nat. Comms.}, {\bf 2016}, \emph{8}, 14567.
\bibitem{AdvMat.30.1707390} Fontanesi, C.; Capua, E.; Paltiel, Y.; Waldeck, D. H.; Naaman, R., Spin-Dependent Processes Measured without a Permanent
Magnet, \emph{Adv. Mater.}, {\bf 2018}, \emph{30}, 1707390.
\bibitem{JPhysChemLett.10.1139} Smolinsky, E. Z. B.; Neubauer, A.; Kumar, A.; Yochelis, S.; Capua, E.; Carmieli, R.; Paltiel, Y.; Naaman, R.; Michaeli, K., Electric Field-Controlled Magnetization in GaAs/AlGaAs Heterostructures-Chiral Organic Molecules Hybrids, \emph{J. Phys. Chem. Lett.}, {\bf 2019}, \emph{10}, 1139.

\bibitem{NatComm.4.2256} Ben Dor, O.; Yochelis, S.; Mathew, S. P.; Naaman, R.; Paltiel, Y., A chiral-based magnetic memory device without a permanent magnet, \emph{Nat. Comms.}, {\bf 2013}, \emph{4}, 2256.
\bibitem{NanoLett.14.6042} Ben Dor, O.; Morali, N.; Yochelis, S.; Baczewski, L. T.; Paltiel, Y., Local Light-Induced Magnetization Using Nanodots and Chiral
Molecules, \emph{Nano Lett.}, {\bf 2014}, \emph{14}, 6042--6049.
\bibitem{Small.15.1804557} Koplovitz, G.; Leitus, G.; Ghosh, S.; Bloom, B. P.; Yochelis, S.; Rotem, D.; Vischio, F.; Striccoli, M.; Fanizza, E.; Naaman, R.; Waldeck, D. H.; Porath, D.; Paltiel, Y., Single Domain 10 nm Ferromagnetism Imprinted on Superparamagnetic Nanoparticles Using Chiral Molecules, \emph{Small}, {\bf 2019}, \emph{15}, 1804557.

\bibitem{AdvMater.28.1957} Kiran, V.; Mathew, S. P.; Cohen, S. R.; Delgado, I. H,; Lacour, J.; Naaman, R., Helicenes—A New Class of Organic Spin Filter, \emph{Adv. Mater}, {\bf 2016}, \emph{28}, 1957--1962.
\bibitem{JPhysChemLett.9.2025} Kettner, M.; Maslyuk, V. V.; N\"urenberg, D.; Seibel, J.; Gutierrez, R.; Cuniberti, G.; Ernst, K. -H.; Zacharias, H., Chirality-Dependent Electron Spin Filtering by Molecular Monolayers of Helicenes, \emph{J. Phys. Chem. Lett.}, {\bf 2018}, \emph{9}, 2025.


\bibitem{JChemPhys.131.014707} Yeganeh, S.; Ratner, M. A.; Medina, E.; Mujica, V., Chiral electron transport: Scattering through helical potentials. \emph{J. Chem. Phys.}, {\bf 2009}, \emph{131}, 014707.
\bibitem{EPL.99.17006} Medina, E.; L\'opez, F.; Ratner, M. A.; Mujica, V., Chiral molecular films as electron polarizers and polarization modulators, \emph{EPL}, {\bf 2012}, \emph{99}, 17006.
\bibitem{JPCM.26.015008} Varela, S.; Medina, E.; L\'opez, F.; Mujica, V., Inelastic electron scattering from a helical potential: transverse polarization and the structure factor in the single scattering approximation, \emph{J. Phys.: Condens. Matter}, {\bf 2013}, \emph{26}, 015008.
\bibitem{PhysRevB.88.165409} Eremko, A. A.; Loktev, V. M.; Spin sensitive electron transmission through helical potentials, \emph{Phys. Rev. B}, {\bf 2013}, \emph{88}, 165409.
\bibitem{JChemPhys.142.194308} Medina, E.; Gonz\'alez-Arraga, L. A.; Finkelstein-Shapiro, D.; Berche, B.; Mujica, V., Continuum model for chiral induced spin selectivity in helical molecules, \emph{J. Chem. Phys.}, {\bf 2015}, \emph{142}, 194308.
\bibitem{PhysRevE.98.052221} D\'iaz, E.; Conteras, A.; Hern\'andez, J.; Dom\'inguez-Adame, F., Effective nonlinear model for electron transport in deformable helical molecules, \emph{Phys. Rev. E}, {\bf 2018}, \emph{98}, 052221.
\bibitem{PhysRevB.99.024418} Yang, X.; van der Wal, C. H.; van Wees, B. J.; Spin-dependent electron transmission model for chiral molecules in mesoscopic devices, \emph{Phys. Rev. B}, {\bf 2019}, \emph{99}, 024418.
\bibitem{NJP.20.043055} D\'iaz, E.; Albares, P.; Est\'evez, P. G.; Cerver\'o, J. M.; Gaul, C.; Diez, E.; Dom\'inguez-Adame, F., Spin dynamics in helical molecules with nonlinear interactions, \emph{New. J. Phys.}, {\bf 2018}, \emph{20}, 043055.
\bibitem{JPhysChemC.123.17043} Michaeli, K.; Naaman, R., Origin of Spin-Dependent Tunneling Through Chiral Molecules, \emph{J. Phys. Chem. C}, {\bf 2019}, \emph{123}, 17043--17048.

\bibitem{PhysRevB.85.081404(R)} Gutierrez, R; D\'iaz, E.; Naaman, R,; Cuniberti, G.; Spin-selective transport through helical molecular systems, \emph{Phys. Rev. B}, {\bf 2012}, \emph{85}, 081404(R).
\bibitem{PhysRevLett.108.218102} Guo, A. -M.; Sun, Q. -F., Spin-Selective Transport of Electrons in DNA Double Helix, \emph{Phys. Rev. Lett.}, {\bf 2012}, \emph{108}, 218102.
\bibitem{JPhysChemC.117.13730} Rai, D.; Galperin, M., Electrically Driven Spin Currents in DNA, \emph{J. Phys. Chem. C}, {\bf 2013}, \emph{117}, 13730--13737.
\bibitem{PhysRevB.93.075407} Matityahu, S.; Utsumi, Y.; Aharony, A.; Entin-Wohlman, O.; Balseiro, C. A., Spin-dependent transport through a chiral molecule in the presence of spin-orbit
interaction and nonunitary effects, \emph{Phys. Rev. B}, {\bf 2016}, \emph{93}, 075407.
\bibitem{PhysRevB.93.155436} Varela, S.; Mujica, V.; Medina, E., Effective spin-orbit couplings in an analytical tight-binding model of DNA:
Spin filtering and chiral spin transport, \emph{Phys. Rev. B}, {\bf 2016}, \emph{93}, 155436.
\bibitem{ChemPhys.477.61} Behnia, S.; Fathizadeh, S.; Akhshani, A., Modeling spin selectivity in charge transfer across the DNA/Gold interface, \emph{Chem. Phys.}, {\bf 2016}, \emph{477}, 61--73.

\bibitem{JPhysChemLett.9.5453} Maslyuk, V. V.; Gutierrez, R.; Dianat, A.; Mujica, V.; Cuniberti, G., Enhanced Magnetoresistance in Chiral Molecular Junctions, \emph{J. Phys. Chem. Lett.}, {\bf 2018}, \emph{9}, 5453--5459.
\bibitem{JPhysChemLett.9.5753} D\'iaz, E.; Dom\'inguez-Adame, F.; Gutierrez, R.; Cuniberti, G.; Mujica, V., Thermal Decoherence and Disorder Effects on Chiral-Induced Spin Selectivity, \emph{J. Phys. Chem Lett}, {\bf 2018},\emph{9}, 5753.
\bibitem{chemrxiv.8325248} Z\"ollner, M. S.; Varela, S.; Medina, E.; Mujica, V.; Herrmann, C., Chiral-Induced Spin Selectivity: A Symmetry Analysis of Electronic Transmission, \emph{unpublished} {\bf 2019}; chemrxiv.8325248 (doi.org/10.26434/chemrxiv.8325248\_v1).



\bibitem{PhysRevLett.61.2015} Kane, C. L.; Mele, E. J., Model for a Quantum Hall Effect without Landau Levels: Condensed-Matter Realization of the "Parity Anomaly", \emph{Phys. Rev. Lett.}, {\bf 1988}, \emph{61}, 2015.
\bibitem{PhysRevLett.95.146802} Kane, C. L.; Mele, E. J., $Z_2$ Topological Order and the Quantum Spin Hall Effect, \emph{Phys. Rev. Lett.}, {\bf 2005}, \emph{95}, 146802.
\bibitem{PhysRevLett.95.226801} Kane, C. L.; Mele, E. J., Quantum Spin Hall Effect in Graphene, \emph{Phys. Rev. Lett.}, {\bf 2005}, \emph{95}, 226801.

\bibitem{PhysRevB.72.075314} Fransson, J., Nonequilibrium theory for a quantum dot with arbitrary on-site correlation strength coupled to leads, \emph{Phys. Rev. B}, {\bf 2005}, \emph{72}, 075314.
\bibitem{NonEquilibriumNanoPhysics} Fransson, J., \emph{Non-Equilibrium Nano-Physics}, Springer: Dordrecht, 2010.

\bibitem{PhysRevB.72.045415} Fransson, J., Angular conductance resonances of quantum dots strongly coupled to noncollinearly oriented ferromagnetic leads, \emph{Phys. Rev. B}, {\bf 2005}, \emph{72}, 045415.

\end{thebibliography}
\end{document}